# Short-range order and its impacts on the BCC NbMoTaW multi-principal element alloy by the machine-learning potential

Pedro A. Santos-Florez[1], Shi-Cheng Dai[2], Yi Yao[3], Howard Yanxon[1], Lin Li[3, *], Yun-Jiang Wang[2, *], Qiang Zhu[1, *], Xiao-Xiang Yu[4, *]

1. Department of Physics and Astronomy, University of Nevada, Las Vegas, NV 89154, USA
2. State Key Laboratory of Nonlinear Mechanics, Institute of Mechanics, Chinese Academy of Sciences, Beijing 100190, China
3. Department of Metallurgical and Materials Engineering, University of Alabama, Tuscaloosa, AL 35401, USA
4. Department of Materials Science and Engineering, Northwestern University, Evanston, IL 60208, USA

**Abstract**

We employ a machine-learning force field, trained by a neural network (NN) with bispectrum coefficients as descriptors, to investigate the short-range order (SRO) influences on the BCC NbMoTaW alloy strengthening mechanism. The NN interatomic potential provides a transferable force field with density functional theory accuracy. This novel NN potential is applied to elucidate the SRO effects on the elasticity, vibrational modes, plasticity, and strength in the NbMoTaW multi-principal element alloy (MPEA). The results show the strong attraction among Mo-Ta pairs forming the local ordered B2 structures, which could be tuned via temperature and improved by Nb content. SRO increases the elastic constants and high-frequency phonon modes as well as introduces extra lattice friction of dislocation motion. This approach enables a rapid compositional screening, paves the way for computation-guided materials design of new MPEAs with better performance, and opens avenues for tuning the mechanical properties by processing optimization.

*Corresponding authors: lin.li@eng.ua.edu, yjwang@imech.ac.cn, qiang.zhu@unlv.edu, and yuxx07@gmail.com

# 1. Introduction

The demand for higher fuel efficiency and lower carbon emission has led to a pressing need to discover, develop, and deploy novel materials that can sustain high mechanical and corrosion damage in extreme environments. After many years of improvements, developing conventional high-temperature materials such as nickel or cobalt-based superalloys, stainless steel, and refractory alloys has plateaued. The emerging multi-principal element alloys (MPEAs) have a vast yet under-explored compositional space, and the numerous combinations of constituents offer considerable freedom in the material design [1-4].

Among a wide range of material properties observed on various compositions and microstructures, some MPEAs have shown exceptional mechanical properties and degradation resistance at elevated temperatures [5-9]. For example, Senkov et al. [10] compared the temperature dependence of the yield stress of NbMoTaW and VNbMoTaW refractory MPEAs and Ni-based superalloys. The results show the strength of Inconel 718 rapidly decreases above 800 °C. In contrast, the yield stress of the two MPEAs gradually decreases from 600 °C to 1600 °C. Moreover, multiple components combined have a stronger resistance to high-temperature softening than the individual refractory element constituent [11].

Currently, the origin of the distinct properties in the BCC MPEAs remains elusive. Among all the MPEAs' characteristics, chemical short-range order (SRO) has been suggested to play a vital role [12]. Compared to conventional alloys [13-16], MPEAs are likely to exhibit more substantial SRO effects due to the multi-principal components and highly concentrated composition. As such, the local ordering can introduce unusual dislocation slip modes and deformation mechanisms [17], such as the comparable velocities of edge and screw dislocations in BCC MPEAs [18], to modify the macroscopic mechanical properties. To verify this hypothesis, it is necessary to conduct a detailed study to characterize the impacts of SRO on the mechanical properties at the atomic level.

Experimental determination of the SRO is exceptionally challenging. Although SRO has been characterized via various scattering technologies (e.g., x-ray, neutron, and electron), more direct observations by advanced electron microscopy became available only recently [19-21]. In terms of the theoretical predictions, people use the cluster expansion approach to expand the alloy's structure into multiple atomic clusters and, through the density functional theory (DFT), calculate energy to determine the interaction coefficient. After that, energy can be calculated for various atomic configurations by Monte-Carlo (MC) sampling [22]. Hybrid DFT and MC method was also employed to sample the multi-component solid solutions [23]. Recently, DFT-based linear response theory and concentration wave analysis were applied to directly predict the SRO and assess the competing long-range order [24].

Despite the efforts mentioned above, the DFT-based methods are difficult to perform large-scale simulations and examine the influence of SRO on the deformation mechanisms. Instead, conventional interatomic potentials were used to study the dislocation structures and mobilities [25]. But this method heavily depends on the availability and reliability of potential models. Meanwhile, several analytic models were developed to provide a generalized prediction [26], but the theory becomes complicated considering the energy contribution beyond the pair interactions.

Recently, machine learning methods have been widely applied in materials modeling. The machine learning potentials (MLPs) are trained by minimizing the cost function to deliberately attune the model to describe the DFT data. The cost of atomistic simulation is orders of magnitude lower than the quantum mechanical simulation, allowing the system to be scaled up to 1 million atoms [27, 28]. To date, several regression techniques, including neural networks (NN) [29], Gaussian process regression [30], and line regression [31, 32], are popular choices for MLP development. Compared to other regression techniques, the NN approach has an unbiased mathematical form that can be adapted to any set of reference points through an iterative fitting process on an extensive training data set. To gain a better predictive power, several types of symmetry-invariant structure descriptors have been developed to represent the local atomic environment that goes beyond the traditional representation in Cartesian coordinates [33]. Many applications based on different MLP models have shown that the machine learning approach works remarkably well in various atomistic simulations [34-36]. In particular, several MLP models have been developed to investigate the MPEA systems [35-38]. These encouraging results promise using MLP to resolve the dilemma of compromising accuracy and cost for traditional models based on DFT or classical force fields.

In this work, we apply the MLP-aided atomistic simulation to understand the role of SRO in strengthening mechanisms on a model NbMoTaW system. In the following sections, we will begin with the introduction of MLP development and other necessary computational approaches employed in this study. The MLP is then applied to different types of simulations (e.g., hybrid molecular dynamics/Monte Carlo, vibrational analysis, and mechanical deformation) to construct the models with and without SROs and map the relation between SRO and the target mechanical properties. Based on the simulation results, we will discuss the interplay between SRO, phase stability, dislocation core structures, plasticity, and strength. Our approaches are general and can be used to investigate other MPEA systems and tune the mechanical properties through structure-composition-processing optimization.

**2. Machine-learning potential development and computational methods**

2.1 Machine-learning potential development

Several software packages have been developed to promote the application of MLPs through different protocols [39-43]. Among them, we have developed the PyXtal_FF package [44] that can train different MLP models via the customized choices of machine learning regressions and descriptors. According to our experience with many other systems [44], we chose the NN regression of Spectral Neighbor Analysis Potential (NN-SNAP). This model has been interfaced with the ML-IAP package in the LAMMPS software [45]. In addition to the available data (5529 configurations from MD simulation and special quasi-random structure modeling) in a recent work [37], we added hundreds of configurations with large elastic strains and different crystallographic directions to improve the model's predictive capability. Finally, we carefully designed the model to deal with two extreme scenarios that might appear under high-temperature MD simulations. The first case is the atomic contact at a short distance due to sizable thermal fluctuation. Such configurations are often far from equilibrium and cannot be interpolated by the training data. Therefore, we added an explicit physical term to express the short-range repulsion through the Ziegler-Biersack-Littmark empirical potential [46]. The second case is that a model with a large size may possess defects where the local atoms have low coordination. Such extreme environments have been largely ignored when people construct the DFT training data set from small size structure periodic models. To address this issue, we also included a set of dimer and trimer configurations (e.g., Mo-Mo, Mo-Ta-Mo with different atomic distances) in our training data to ensure that the trained model understands such distinct environments.

With the augmented data, we performed DFT calculations using the Perdew-Burke-Ernzerhof functional implemented in the VASP code [47], with a plane-wave cutoff energy of 520 eV. For each structure, we computed the SNAP descriptor for each configuration with a band limit ($2j_{max}$) up to 6, corresponding to 30 bispectrum components. The NN training was executed with two hidden layers with 30 nodes for each layer while energy, force, and stress contributions were trained simultaneously. The importance coefficients of force and stress were set to $10^{-3}$ and $10^{-4}$, respectively. The results of the NN potential training are illustrated in Figure 1. Overall, the mean absolute errors are 4.403 meV/atom for energy, 0.1 eV/A for force, and 0.532 GPa for stress tensors. The MLP accuracy is comparable with other recent works [35-37], thus warranting its application for the studies to be discussed in the following sections.

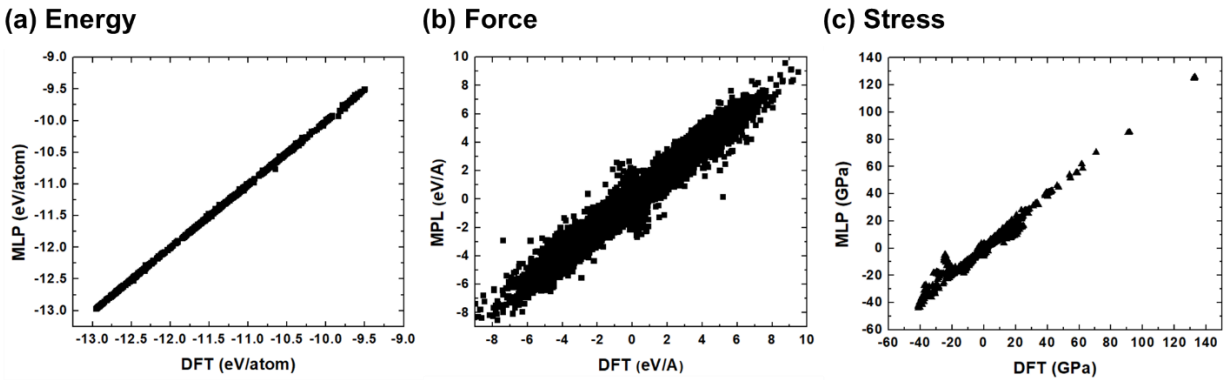

Figure 1. The performance of (a) energy, (b) force, and (c) stress tensor training.

2.2 Hybrid MD/MC simulations

The local chemical orders in BCC MoNbTaW solid-solution alloy were investigated by the LAMMPS software using NN-SNAP. MC swaps of atoms at the specified temperature were performed with the acceptance probability based on the Metropolis criterion [48]. All atoms in the simulation domain were moved using time integration on Nose-Hoover style and sampled from the canonical ensembles, resulting in a hybrid MC/MD simulation.

2.3 Mechanical property simulations

2.3.1 Elastic constants
In order to understand the elastic behavior under small deformations and the SRO effect, the elastic constants were calculated for samples of BCC MoNbTaW alloys after hybrid MD/MC at 300 K and 1200 K. In the first step, the samples containing 8192 atoms were equilibrated with the NPT (isothermal-isobaric) ensemble at zero pressure and a finite temperature varying from 100 K to 1200 K for approximately 15 ps. The MD equilibrations were stopped when the fluctuating pressure was closer to zero. In the second step, a negative and a positive strain of 1% was applied separately to the computational cells for each of the Voigt deformation directions (xx→1, yy→2,

zz→3, yz→4, xz→5, xy→6). In the final step, the deformed computational cells were equilibrated with the NVT (canonical) ensemble at the same initial temperature for 10 ps. At 0 K, conjugate gradient energy minimizations were performed instead of MD equilibrations.

The resultant average changes in stress were used to compute one row of the elastic stiffness tensor for each Voigt deformation component, considering the average of the negative and positive deformations. Therefore, the elastic constants were calculated using the cubic symmetries as

$C_{11} = (C_{11} + C_{22} + C_{33})/3$ \hfill (1)
$C_{12} = (C_{12} + C_{21} + C_{13} + C_{31} + C_{23} + C_{32})/6$ \hfill (2)
$C_{44} = (C_{44} + C_{55} + C_{66})/3$ \hfill (3)

2.3.2 Phonon density of states

To understand the elastic stability and SRO's influence on the vibrational properties, we calculated the phonon density of states (PDOS) of alloy samples with different degrees of SRO, which were introduced by thermal annealing via the MC/MD algorithm. The phonon DOS was estimated using the Fourier transform of the velocity-velocity autocorrelation function (VACF) derived from the MD trajectories [49]. This standard procedure is formulated as

$$g(\omega) = \int_0^\infty e^{i\omega t} \frac{\langle \vec{v}(t)\vec{v}(0) \rangle}{\langle \vec{v}(0)\vec{v}(0) \rangle} dt, \hfill (4)$$

where $\omega$ is the vibrational angular frequency of a normal mode and $g(\omega)$ is the phonon DOS. $\vec{v}(t)$ is the velocity of an atom at time $t$, and $\vec{v}(0)$ is the initial velocity. Therefore, the VACF $\langle \vec{v}(t)\vec{v}(0) \rangle$ can be obtained directly from the instant velocity of each atom from the output of any molecular dynamics code. $\langle \cdots \rangle$ denotes an ensemble average. The integral goes sufficiently long to guarantee all the vibrational modes are included. The time resolution should be high enough in MD to avoid statistical error and huge fluctuation in DOS. Furthermore, the simulation box should be large enough to include the long wavelength vibrational mode with low frequency. Therefore, our simulation model had 520,000 atoms with $64 \times 64 \times 64$ BCC lattices, which is believed to be large enough to produce physically sound VACF and reliable phonon features [50].

2.3.3 Generalized stacking fault

Generalized stacking fault (GSF) energies are essential parameters that influence dislocation mobility, deformation twinning, and phase transformation. Here, the GSF energies of the primary slip system <111>{110} in BCC MoNbTaW were performed using a large supercell containing about 36,000 atoms with and without SRO. The supercell was set to be periodic along [111] and [11-2] directions and non-periodic along [-110].

2.3.4 Peierls stress

To estimate the role of SRO plays in strength, we computed the Peierls stress of a screw dislocation in both ordered and disordered samples. Peierls stress is the threshold stress when dislocation starts to glide at 0 K in an infinite crystal without other defects. It indicates the lattice friction of dislocation, thus a baseline for the strength of metals. To calculate the Peierls stress, an athermal quasistatic shear deformation was applied with a small increment of strain $10^{-4}$ until to a critical point where the dislocation started to move. At each strain, the model was fully relaxed to its local energy minimum.

2.3.5 Tensile strain-stress curve

Uniaxial deformations were also performed. Prior to deformation, the samples were equilibrated at zero pressure and the desired temperature with the NPT ensemble for 10 ps. After the initial equilibration, uniaxial tensile deformations were loaded along the [001], [110] and [111] crystallographic directions with a strain rate of $1 \times 10^8$ s$^{-1}$. The stress was measured along the deformation direction, maintaining zero pressure in the transverse directions at constant temperature. For deformations along the [001] direction, the simulations were carried out using a periodic conventional orthogonal cell containing 8192 atoms with $16 \times 16 \times 16$ BCC lattices. On the other hand, for the deformations along the [110] and [111] directions, a rotated computational cell containing 7128 atoms was used, changing the conventional orthogonal orientations by ([1,1,1], [1,1,-2], [-1,1,0]).

## 3. Results

3.1 Chemical SRO and its temperature dependence in MoNbTaW

To quantify the chemical ordering around a specific element, we calculated the Warren–Cowley (W-C) parameters of the atomic configurations after hybrid MD/MC sampling at different temperatures using the following equation [51]:

$$\alpha_{ij}^k = 1 - p_{ij}^k / c_j \qquad (5)$$

where $p_{ij}^k$ is the probability of finding atom $i$ in the $k$th coordination shell surrounding the atom $j$. $c_j$ is the nominal concentration of element $j$. In the case of $\alpha_{ij}^k < 0$, it suggests attractive interactions between the two atom types while an $\alpha_{ij}^k > 0$ suggests repulsion.

From the results in Figure 2(a) for various first nearest-neighbors, Mo-Ta, Ta-W, Mo-Nb, and Nb-W pairs show attraction, among others, while Mo-Mo, Ta-Ta, Nb-Ta, and Mo-W pairs show repulsion. Moreover, the SRO is temperature dependent with a disorder-order transition temperature at approximately 1100 K. Above 1100 K, the alloy tends to a random solid solution, and the SRO becomes more potent at lower temperatures below 1100 K. Figure 2(b) compares the equilibrium atomic configurations at 300 K and 1200 K respectively, the squares label the ordered B2 structures formed by Mo-Ta pairs. The fraction of short-range B2 order increases as the temperature decreases, which is consistent with our W-C parameters predicted and prior results reported [22, 36, 37].

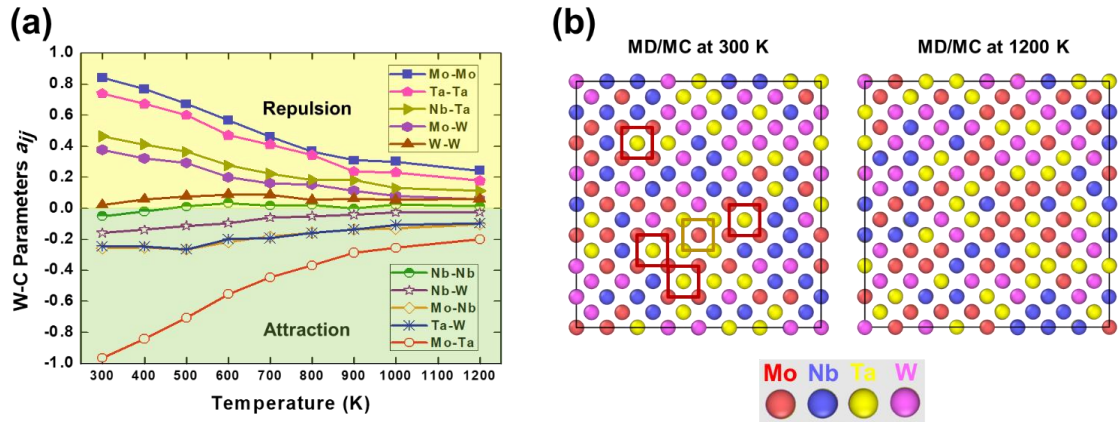

Figure 2. (a) W-C parameters versus the temperatures in MoNbTaW alloy. (b) Atomic configurations at 300 K and 1200 K after MD/MC sampling, respectively, the red and gold squares label the ordered B2 structures formed by Mo-Ta pairs.

3.2 Element effects on SRO in non-equiatomic alloys

To investigate the alloy composition's impact and multiple elements' synergy effects on the SRO, we constructed several non-equiatomic MoTaNbW quaternaries to evaluate the variation in the local chemical environments. More specifically, we focus on the influence of Nb and W on the Mo-Ta pairs and their ordered B2 structures. Both Mo and Ta concentrations were fixed at 25 at.%, and the Nb and W concentrations varied, constructing two more alloy compositions with $Mo_{25}Ta_{25}Nb_xW_{50-x}$ (x=10 and 40). Bulk BCC supercells of 8192 atoms were annealed via MD/MC simulations with the ML potential at 300 K and 1200 K, representing SRO and random structures, respectively.

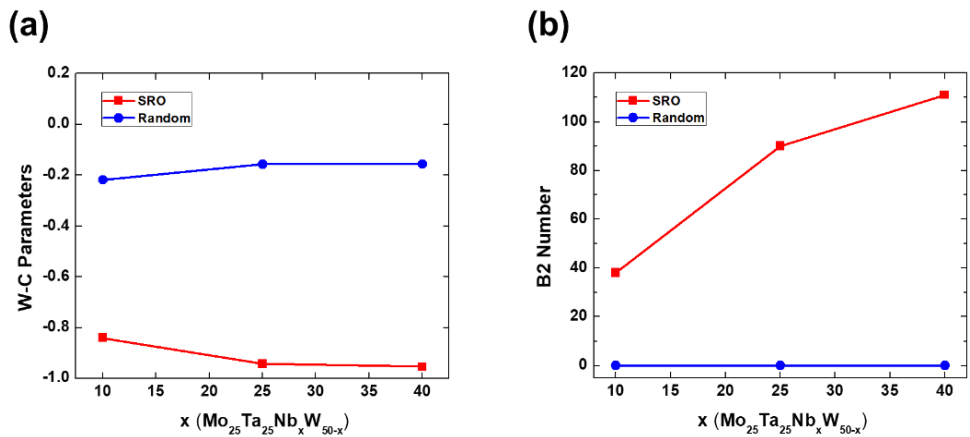

Figure 3. (a) W-C parameters of the Mo-Ta pair in the SRO and random MoTaNbW quaternaries. (b) The number of Mo-Ta B2 structures in the three alloy compositions.

Figure 3 presents the W-C parameters of the Ta-Mo pair in the two alloy variations, along with the equiatomic MoTaNbW for comparison. The W-C parameters of the Mo-Ta pair decrease with Nb content in the SRO structures equilibrated at 300 K, indicating Nb enhances the ordering of Mo and Ta. We further counted the number of Mo-Ta B2 structures in all the alloys, where eight first nearest Mo neighbors surrounded one centered Ta atom and vice versa. Notably, the Nb element monotonically increases the Mo-Ta B2 structures among the three compositions, even though the Mo and Ta concentrations remain unchanged. For instance, $Mo_{25}TaNb_{40}W_{10}$ has 111 Mo-Ta B2 structures out of 8192 atoms in the BCC supercell, which is about 23% more than 90 B2 structures in the equiatomic MoTaNbW. On the other hand, W exhibits a negative effect on forming the Mo-Ta B2 structure.

In the subsequent investigation of SRO effects, we chose the atomic configurations after MD/MC at 300 K and 1200 K as representative ordered (with SRO) and disordered (without SRO) models and calculated several mechanical properties for comparison. In the following discussions, these two types of samples will be referred to as SRO and Random, respectively.

3.3 Elastic constants

Figure 4 shows the elastic constants of the BCC MoNbTaW alloy as a function of temperature obtained from the SRO and random samples. The results show an approximately linear relationship of the elastic constants with respect to temperature. As expected, the alloys are more compliant at higher temperatures. $C_{11}$ is more sensitive to temperature, which corresponds to uniaxial deformations. The $C_{11}$ and $C_{12}$ of the samples with SRO are about 1% higher than those of random ones. SRO increases the $C_{44}$ by 3.5%~4.5%, introducing more resistance to the shear deformation.

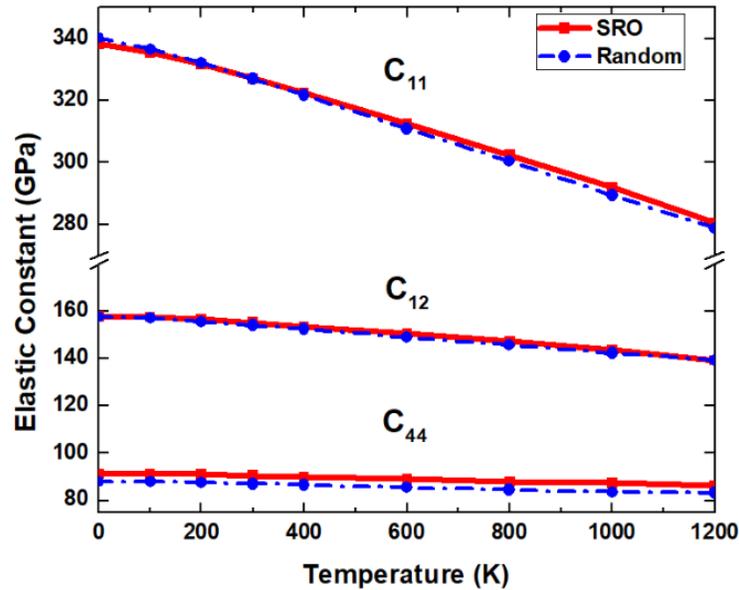

Figure 4. Elastic constants of BCC MoNbTaW alloy at different temperatures were obtained from MD/MC simulations for both SRO and random samples.

3.4 Phonon density of states

Figure 5(a) shows the velocity-velocity autocorrelation functions (VACFs) of specific alloys at the temperature of 10 K. The VACFs decay very quickly and fluctuate around zero within a few picoseconds. The phonon density of states (PDOS) can be obtained from this feature by Fourier transform according to Eq. (4) and illustrated in Fig. 5(b). Several characteristics can be noticed from the vibrational properties. First, the MD calculations are reliable if one compares the DOS of a random sample to the DFT data of random alloys obtained using a special quasi-random structure supercell model [52]. At very high frequencies above 30 meV, DOS estimated from MD vanishes, but there are still some high-frequency modes from the DFT prediction. Next, any alloy has no imaginary vibrational frequency, indicating they are thermodynamically and elastically stable. Third, SRO leads to increased PDOS and enhancement in high-frequency modes beyond 25 meV. The low-frequency modes (mainly corresponding to the elastic waves) remain nearly unchanged, consistent with our aforementioned results that SRO does not remarkably increase elastic constants. However, the high-frequency modes change significantly, implying that the SRO stiffening is only a local effect [53, 54]. Finally, the enhanced local hardening may raise extra resistance to the dislocation motion, as shown in the following Peierls stress results.

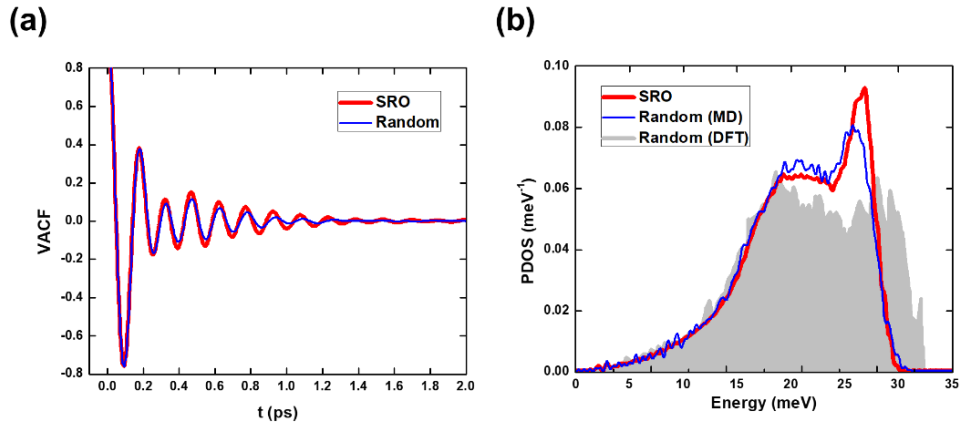

Figure 5. Vibrational features of BCC MoNbTaW alloy. (a) The VACFs were obtained from MD simulations for samples with the SRO and random ones. (b) Phonon DOS of alloys with different SROs. Normal modes move to high frequency with increasing chemical SRO. MD calculation with the present MLP (solid blue curve) is validated by DFT calculations (gray area) on random structures [52].

3.5 GSF energies

GSF energies were calculated on five {110} planes along with the slip direction <111> as shown in Figure 6(a). The averaged GSF in Fig. 6(b) shows that the unstable stacking fault energy (maxima on the GSF energy curve) increases by ~100 mJ/m$^2$ in the SRO model compared to the random one. The error bars in the energy curves for each displacement indicate the impacts of chemical fluctuation at various locations in the multiple components and highly concentrated alloy; however, the variation is less than 20 mJ/m$^2$. Therefore, the significant contribution to the GSF energy increment and resulting resistance to dislocation slip and dissociation is due to the break of SRO bonding. As an example shown in Figure 6(a), the Mo-Ta B2 structures were broken on the (110) slip plane during the shift of two crystal halves relative to each other.

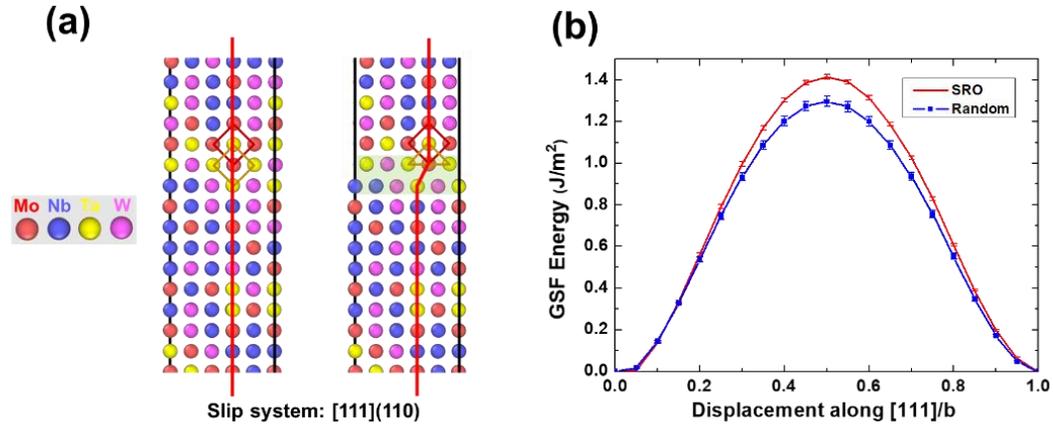

Figure 6. (a) The side view images of atomic configurations in the original SRO structure and stacking fault. The red and gold squares label the ordered Mo-Ta B2 structures. (b) Averaged GSF of MoNbTaW with and without SRO. The error bars denote chemical fluctuation at various locations.

3.6 Peierls stress

For Peierls stress simulation, a $\frac{1}{2}\langle 111 \rangle$ screw dislocation was modeled in the supercell of Figure 7(a), and the burger vector was along the *x* [111] direction. The strain-stress curves are shown in Fig. 7(b). Notice that there are a series of stress fluctuations around strain 0.02 in the random structure (blue line). This fluctuation is due to the local structural relaxation rather than Peierls stress. The calculated Peierls stresses in MoNbTaW are 2.47 GPa, and 2.91 GPa, for the random and SRO structures, respectively. Therefore the SRO increases both the lattice friction of dislocation motion and the strength of alloy, which are consistent with the enhancement in high-frequency vibrational modes and elastic moduli due to SRO.

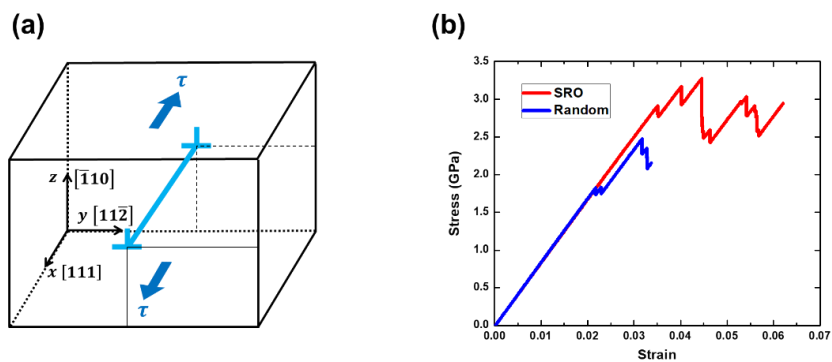

Figure 7. Resistance of dislocation mobility in MoNbTaW alloys. (a) Schematic illustration of the simulation cell geometry for assessment of Peierls stress of a screw dislocation. (b) Shear stress-strain curves for the SRO and random sample.

3.7 Tensile behaviors

To investigate the deformation mechanisms under external strains, we perform a tensile loading at a strain rate of $10^8$/s along the [100], [110], and [111] directions for both the SRO and random samples. Figure 8 shows the stress-strain response is anisotropic, depending on crystallographic orientations. The SRO slightly increases the modulus (slope) in the elastic deformation stage for all three directions, consistent with our previous results. SRO and random samples exhibit similar yielding stress. Still, the SRO sample shows more deformation resistance in the [100] and [110], as evidenced by the higher stress level to keep deforming after yield.

Figure 9 displays several representative atomic configurations from the deformation paths on the SRO sample to understand the mechanisms. Fig. 9 (a) presents the [100] tensile stress-strain curve, along with several critical moments denoted on it. No imperfections are found in the initial sample before the start of the tensile process. It is noted that the mechanical instability occurs around a strain of 11% and stress of 16 GPa, driving the simulated sample away from the elastic limit. As the stress increases, <112>(111) twinning is observed in the crystalline lattice (ref. to Fig. 9 Strain=0.13). The first significant stress drop to a lower value ~ 6.2 GPa, followed by a serration flow with slightly hardening. Such stress serration is associated with twin boundary migration (ref to Fig. 9 Strain=0.16). At a strain of 0.2, the second major stress drop occurs, accompanied by forming a set of cross-twins (ref. to Fig. 9 Strain=0.21). Very few dislocation activities are captured. One is shown in Fig. 9 (Strain=0.21), where a dislocation nucleates from one twin boundary, propagates across and is absorbed by another twin boundary.

In comparison, the dominant mechanisms in [110] and [111] directions comply with the dislocation-mediated deformation, as shown in Fig.9 (b) and (c). The majority dislocation type is ½<111> with minor <100> type. The following work hardening after yielding is due to the interplay among the dislocations.

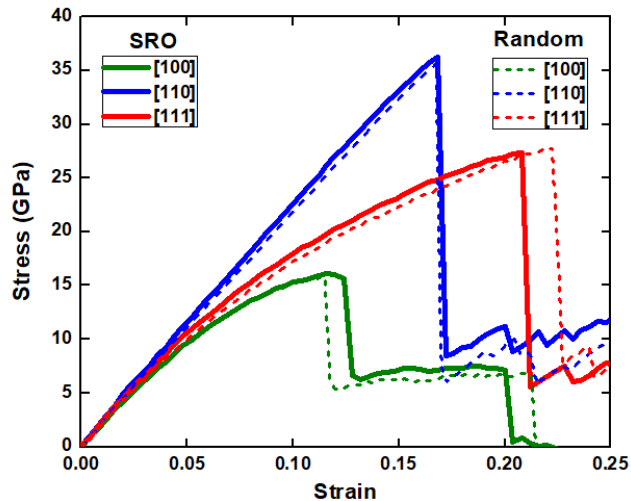

Figure 8. Uniaxial tensile stress-strain response of the SRO (solid line) and random NbMoTaW (dashed line) samples along [100], [110], and [111] directions at 300 K.

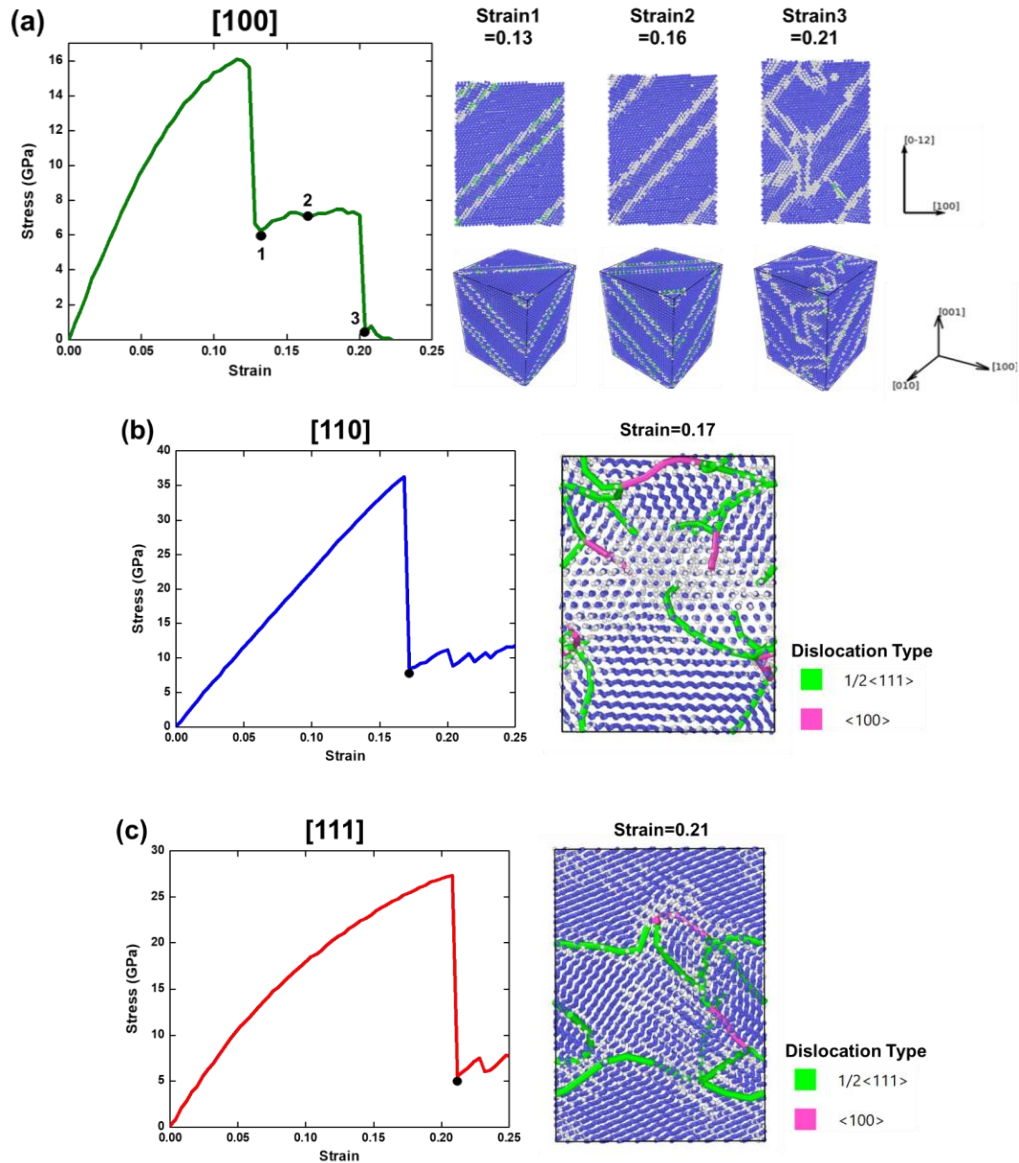

Figure 9. Atomic configurations of sample at various tensile strains along (a) [100], (b) [110], and (c) [111] directions.

## 4. Discussion

Compared to pure BCC metals and dilute alloys, the strength of some refractory MPEAs exhibits weak temperature dependence over a wide temperature range [10]. Currently, the origin of the outstanding properties doesn't have a clear mechanistic picture. One possible reason is that the highly concentrated solution and local chemical fluctuation lead to a rugged energy landscape and considerable variation in barriers of dislocation motion. Another mechanism is through the prevailed SRO in MPEAs, which impacts the structure and mobility of dislocations and enhances the mechanical property. However, the dominant factor in the BCC MPEAs strengthening has yet to be elucidated.

This study scrutinizes the temperature and composition dependence of SRO and their effects on the elastic constants and stability, GSF, Peierls stress, and tensile deformation in the MoNbTaW alloys to deconvolute the complexity of strengthening in MPEAs into individual mechanisms. The results show the strong attraction among Mo-Ta pairs forming the short-range ordered B2 structures. While SRO moderately increases the elastic constants, it evidently results in the redistribution of high-frequency phonon modes, which is critical to the material's thermodynamic properties at elevated temperatures.

The remarkable improvement of GSF and Peierls stress indicates extra lattice friction of dislocation motion by SRO besides the highly concentrated solution introduced in the rough energy landscape. SRO does not change the twinning initiation in our current tensile deformation compared to the random structure. But it could be expected that the SRO would influence the dislocation structure and velocity after the twin formation.

Rich in SROs is one of the salient features inherent to MPEAs, distinguishing them from traditional alloys. Experimentally, the MPEAs are processed, homogenized, and annealed at temperatures below their melting points, unavoidably accommodating SROs [4]. Tuning the SROs could offer a new strategy to tailor novel multi-component alloys' structural, chemical, magnetic, and mechanical properties. In CrCoNi FCC alloys, it has been shown that tempering to promote SRO has increased hardness, enhanced SFE, and a subsequent increase in planar slip [20]. Similarly, our temperature dependence results map out the tunable SRO degree in MoNbTaW with temper, which facilitates processing optimization to achieve the alloy's applicable property.

In addition to thermal processing, we demonstrate that properly designing the alloy composition by increasing Nb content can enhance the chemical SRO and Mo-Ta ordered B2 units in the non-equiatomic MoTaNbW alloys. Considering that Nb is the lightest element in the MoTaNbW system and further promotes the SRO effect, it is reasonable to anticipate an enhanced specific strength, which provides a guide to the experimentalists in the composition search of this refractory alloy for lightweight applications.

Finally, all the aforementioned simulations heavily rely on the availability of the MLP model. Our MLP model is similar to the previous efforts based on linear regression [35, 36]. However, we believe that using neural networks makes the model more flexible in capturing target functions' nonlinear behavior, such as the potential energy surface. More importantly, the execution cost of NN is similar to linear regression when the same set of descriptors is used. Therefore, it is perhaps more suitable for large-scale simulation of multiple-component systems in which nonlinear effects are more pronounced.

Moreover, it is essential to note that nearly all MLP models suffer from the extrapolation problem. Our initial MLP model correctly predicted all elastic and vibrational properties and the deformation responses in the [100] direction. However, it yielded an unphysical amorphization behavior for the deformation at [110] and [111]. To improve the MLP, we intentionally included more deformation configurations in our training database. This example highlights the extrapolation challenge in MLP development. One needs to ensure that the simulation does not go beyond the training domain for various application cases for different application cases. Otherwise, the training database must be fed additional data to guarantee the model's interpolative capability.

Only when the training data has been efficiently sampled the MLP-based interpolation can be considered a compelling solution to bridge the gap between DFT and classical force field simulations.

## 5. Conclusion

In summary, we report the application of well-trained MLP to simulate the model MPEA system of NbMoTaW. We trained a neural network potential based on a large set of structure-properties data and MD configurations from DFT calculation. Combining the MLP with hybrid molecular dynamics/Monte Carlo simulations, we thoroughly investigate the impacts of SRO on phase stability, dislocation core structures, plasticity, and strength in the NbMoTaW MPEA. In addition to the NbMoTaW with equal proportions, the combination of MLP with other contemporary computational techniques is also suitable for rapid screening of target compounds in a vast compositional space, thus paving the way for computation-guided materials design of new MPEAs with better performance.

## Acknowledgments

The computing resources are provided by XSEDE (TG-DMR180040). P. A. S., H. Y., and Q. Z. thank Dr. Aidan Thompson for helpful discussions regarding implementing NN-SNAP into the LAMMPS package. Y.J.W was supported by the NSFC (Grant No. 12072344) and the Youth Innovation Promotion Association of the Chinese Academy of Sciences.